\documentclass[aip,apl,reprint,noshowpacs,noshowkeys,amssymb,amsmath,amsfonts,bibnotes]{revtex4-2}

\makeatletter
\let\@fnsymbol\@fnsymbol@latex
\@booleanfalse\altaffilletter@sw
\def\frontmatter@makefnmark{\@textsuperscript{\normalfont\@thefnmark}}
\makeatother
\usepackage{physics}
\usepackage{graphicx}
\usepackage{theorem}
\usepackage{dcolumn}
\usepackage{longtable}
\usepackage{bm}
\usepackage{ulem}
\usepackage{accents}
\usepackage{color} 
\usepackage[dvipsnames]{xcolor}

\usepackage{makecell}

\renewcommand{~}{\,}

\newcommand{\SSS}{\scriptscriptstyle}

\newcommand{\ie}{{\textit{i.e.}}}

\newcommand{\Ee}{{\rm e}}
\newcommand{\Dd}{{\rm d}}

\newcommand{\Ii}{{\rm i}}

\newcommand{\cNe}{c_{\SSS\text{Ne}}}
\newcommand{\rhoNe}{\rho_{\SSS\text{Ne}}}

\newcommand{\epsNe}{\epsilon_{\SSS\text{Ne}}}
\newcommand{\rpar}{\bm{r}_{\SSS\parallel}}

\newcommand{\qpar}{\bm{q}_{\SSS\parallel}}
\newcommand{\qmpar}{q_{\SSS\parallel}}

\newcommand{\oppar}{\hat{\bm{p}}_{\SSS\parallel}}

\newcommand{\rB}{r_{\SSS\text{B}}}
\newcommand{\KB}{k_{\SSS\text{B}}}

\renewcommand{\thetable}{\arabic{table}}
\makeatletter
\renewcommand{\figurename}{Fig.}
\renewcommand{\tablename}{Table}
\renewcommand{\fnum@figure}[1]{\textbf{\figurename~\thefigure}}
\renewcommand{\fnum@table}[1]{\textbf{\tablename~\thetable} }
\makeatother

\makeatletter
\renewcommand{\figurename}{Fig.}
\renewcommand{\tablename}{Table}
\renewcommand{\fnum@figure}[1]{\textbf{\figurename~\thefigure.} }
\renewcommand{\fnum@table}[1]{\textbf{\tablename~\thetable.} }
\makeatother

\begin{document}

\title{Electron charge coherence on a solid neon surface}

\author{Xinhao Li}
\email{xinhaoli@fas.harvard.edu}
\affiliation{Department of Physics, Harvard University, Cambridge, Massachusetts 02138, USA \looseness=-1}

\author{Shan Zou}
\email{szou3@nd.edu}
\affiliation{Department of Physics and Astronomy, University of Notre Dame, Notre Dame, Indiana 46556, USA \looseness=-1}

\author{Qianfan Chen}
\affiliation{Center for Nanoscale Materials, Argonne National Laboratory, Argonne, Illinois 60439, USA \looseness=-1}

\author{Dafei Jin}
\affiliation{Department of Physics and Astronomy, University of Notre Dame, Notre Dame, Indiana 46556, USA \looseness=-1}
\affiliation{Center for Nanoscale Materials, Argonne National Laboratory, Argonne, Illinois 60439, USA \looseness=-1}

\date{\today}

\begin{abstract}
Recent experiments show $\sim$0.1~ms coherence time for a single electron charge qubit on a solid neon surface. This remarkably long coherence time is believed to result from the intrinsic purity of solid neon as a qubit host. In this paper, we present theoretical studies on the decoherence mechanisms of an electron's charge (lateral motional) states on solid neon. At the typical experimental temperature of $\sim$10~mK, the two main decoherence mechanisms are the phonon-induced displacement of neon surface and phonon-induced modulation of neon permittivity (dielectric constant). With a qubit frequency increasing from 1~GHz to 10~GHz, the charge coherence time decreases from about 366~s to 7~ms and from about 27~s to 0.3~ms, respectively, limited by the two mechanisms above. The calculated coherence times are at least one order longer than the observed ones at $\sim$6.4~GHz qubit frequency, suggesting plenty of room for experimental improvement. 
\end{abstract}

\maketitle
\pretolerance=8000 


\section{\label{sec:level1}Introduction}

Among solid-state qubits~\cite{chatterjee2021,siddiqi2021engineering}, electron charge qubits offer promising prospects to serve as building blocks in quantum computing, due to their inherent simplicity in design~\cite{veldhorst2014, takeda2016}, fabrication~\cite{osman2021}, control~\cite{pashkin2009, diVincenzo2016, Kroll2019}, and readout \cite{loss2022, takeda2016, blais2021, burkard2019, nori2012}. However, a notable drawback of electron charge qubits is their generally short coherence times, attributed to their sensitivity to environmental charge noises. The pursuit of novel electron charge qubits with extended coherence times is highly attractive in quantum information science and technology.

Helium (He) and neon (Ne), noble-gas elements with extreme chemical inertia, spontaneously condense into ultrapure quantum liquids and solids at low temperature and pressure~\cite{guo2025quantum}. Compared with classical liquids and solids, condensed He and Ne have negligible charged excitations or spinful isotopes ($^3$He and $^{21}$Ne). They offer a vacuum-like environment with minimal charge and spin noises for electron qubits~\cite{zavyalov2005, guo2025quantum}. When an excess electron approaches a liquid He or solid Ne surface from vacuum, it can form surface states by two effects. First, a repulsive barrier of approximately 1~eV resulting from the Pauli exclusion between the excess electron and the shell electrons in He and Ne atoms. Second, there is an attractive well from the polarization-induced image charge within the liquid or solid~\cite{cole1969,cole1971,leiderer1992}. A surface electron can be trapped in a quantum-dot structure by on-chip electrodes and coupled with microwave photons in a superconducting quantum circuit~\cite{koolstra2019, koolstra2025high, yang2016}.

In the past few years, we successfully realized the electron-on-solid-neon (eNe) motional (charge) qubits~\cite{zhou2022} and obtained charge coherence times up to 0.1~ms~\cite{zhou2024electron}. Despite already exceeding the coherence times of traditional charge qubits by several orders~\cite{loss2022,heinrich2021}, the current eNe charge coherence times are probably limited by the imperfect surface and redundant electrons trapped on it~\cite{kanai2024single, zheng2025surface}. With optimized growth of solid Ne and a refined electron loading procedure, the eNe charge coherence times may be significantly improved.

\begin{figure}[htb]
	\begin{center}
		\includegraphics[scale=0.14]{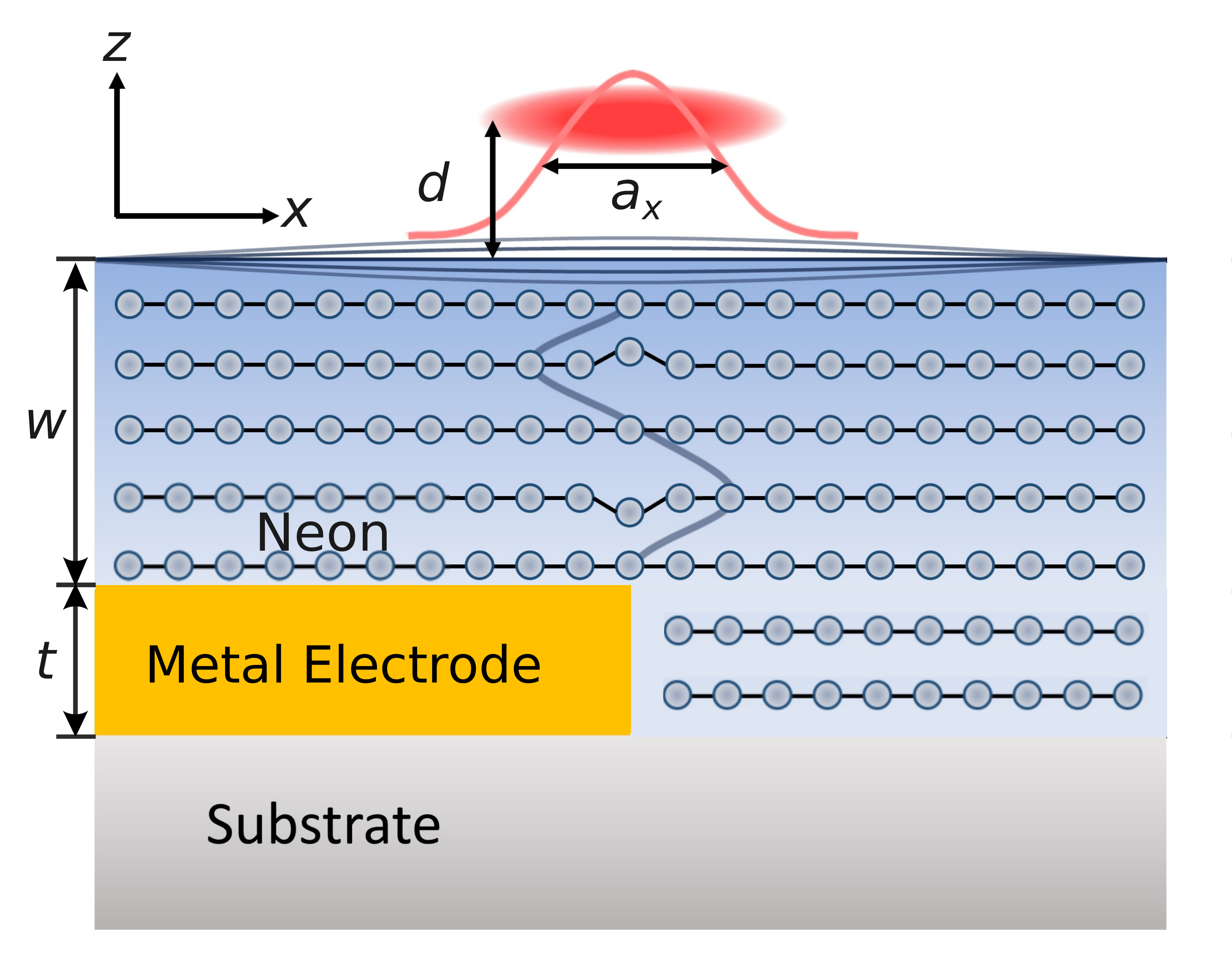}
	\end{center}
	\caption{Schematics of the electron-on-solid-neon (eNe) system, featuring an electron floating in the vacuum above a solid Ne film and electron-phonon interaction. The solid neon film can cover both the metal electrodes and the dielectric substrate of the quantum circuit device. The longitudinal (acoustic) phonons in the bulk induces displacement on the Ne surface. The electron's wavefunction, represented by a red cloud, resides at a mean distance of $d\approx2$~nm from the surface and extends along the surface over a scale of 100~nm. The diagram shows a harmonic ground-state wavefunction (pink curve) with a Gaussian width $a_{x}$.}
	\label{Fig:1-geometry}
\end{figure}

In this paper, we present a theoretical investigation of the intrinsic decoherence mechanisms of the motional (charge) states of a single electron on a perfect solid Ne surface. We calculate the corresponding relaxation and coherence times, which provide the upper limit of the eNe charge coherence times in this system. The system is illustrated in Fig.~\ref{Fig:1-geometry}, which mimics the thin-film structures of typical fabricated devices. A single electron hovers above the surface of a solid Ne film with a thickness denoted by $w$. The devices are typically superconducting circuits made of aluminum (Al) or niobium (Nb) film with a thickness denoted by $t$, deposited on low-microwave-loss substrates such as intrinsic silicon (Si) or sapphire (Al$_2$O$_3$). The solid Ne layer can be grown on the metal electrode and the bare substrate.

\section{Surface charge states}

In this paper, we adopt the simplest model to describe the surface charge states~\cite{andrei2012two}. We approximate the Pauli barrier to be infinitely high, which forces the electron wavefunction to vanish inside solid Ne, $z\leq 0$. We also assume the solid Ne to be thick, $w\gtrsim100$~nm, so the influence from the metal electrode and substrate on the wavefunction is negligible. We first discuss the vertical bound states. The image charge potential in the vertical ($z$) direction is 
\begin{equation}
	V_{\text{img}}(z) = -\frac{\mathit{\Lambda}}{z}, \quad (z>0),
\end{equation}
where $\mathit{\Lambda}$ is the effective Coulomb interaction parameter,
\begin{equation}
	\mathit{\Lambda} = \frac{e^2}{4} \frac{\epsNe-1}{\epsNe+1},
\end{equation}
with $e$ being the elementary charge and $\epsNe=1.244$ being the dielectric constant of uniform solid neon~\cite{cole1969}. 

The vertical state eigenenergies are hydrogenic, 
\begin{equation}
	E_n = -\frac{R}{n^2},\quad n = 1, 2, 3, \dots
\end{equation}
with the effective Rydberg energy,
\begin{equation}
	R =  \frac{\hbar^2}{2m_e\rB^2} = 1.611\times 10^{-14}~\text{erg} = 116.7~\text{K}.
\end{equation}
in which $m_e$ is the electron mass and $\rB$ is the effective Bohr radius,
\begin{equation}
	\rB = \frac{\hbar^2}{\mathit{\Lambda} m_e} = 1.947~\text{nm},
\end{equation}
They are related by $\mathit{\Lambda} = 2R\rB$. The ground-state wavefunction $\left|1_{z}\right\rangle$ is given by
\begin{equation}
	\psi_1(z) = \langle \bm r |1_z\rangle = \frac{2}{\sqrt{\rB}} \frac{z}{\rB} \Ee^{-z/\rB}. \label{Eqn:RydbergGround}
\end{equation}

At the typical experimental temperature, $T=10$~mK, the electron's vertical motion perpendicular to the Ne surface is effectively frozen in the ground state, whereas its lateral motion parallel to the $xy$ plane is confined by the on-chip electrodes. We approximate the trapping potential in each of the lateral dimensions to be harmonic, with level spacing $\hbar\omega_{x,y}$. 
Thus, the Hamiltonian $\hat{\mathcal{H}}_{\text{el}}$ for a single electron is given by
\begin{equation}
	\hat{\mathcal{H}}_{\text{el}} = -\frac{\hbar^2}{2m_e}\nabla^2 + V_{x}(x) + V_{y}(y) + V_z(z),
\end{equation}
with
\begin{equation}
	V_z(z) = V_{\text{img}}(z) + e\mathcal{E}_z z, \quad z>0,
\end{equation}
where $\mathcal{E}_z$ is an electric field that presses the electron against the Ne surface, and 
\begin{equation}
	V_{x}(x) = \frac{1}{2}m_e\omega_x^2 x^2, \quad V_y(y) = \frac{1}{2}m_e\omega_y^2 y^2,
\end{equation}
are lateral harmonic potentials in $x$ and $y$, respectively.

The in-plane motional states of the electron are denoted as $\left\vert n_{x},n_{y}\right\rangle $ with $n_x,n_y = 0, 1, 2,\dots$ being the quantum numbers of quantum harmonic oscillators in $x$ and $y$, respectively. Although in this model an electron's wavefunction entirely resides in the vacuum, phonon excitations in solid Ne can still interact with the electron's motional states, playing the role of a decoherence source, as discussed below.  
\section{Decoherence via photons}

Before discussing phonon-induced decoherece, we first discuss photon-induced spontaneous emission, which sets the upper bound of the relaxation and coherence times in this system.

\subsection{Spontaneous emission in a vacuum}
The electric dipole transition can radiate photons into the surrounding vacuum. The spontaneous emission rate can be described as~\cite{bethe2018}
\begin{equation}
	\mathit{\Gamma} ^{\left( \text{vac}\right) }=\frac{\left\vert d_{x}\right\vert ^{2}%
	}{3 \hbar c^{3}}\omega _{x}^{3}\equiv \frac{1}{%
		T_{1}^{\left( \text{vac}\right) }}.  \label{spontaneous emission rate}
\end{equation}
Here, $d_{x}=\sqrt{2} ea_{x}$ is the electric dipole moment in $x$ with the confining length $a_{x}=\left( \hbar /m\omega_{x}\right) ^{1/2}$, and $f_{x}=\omega_{x}/2\pi$ is the transition frequency. The rate of radiative decay, as given by Eq.~(\ref{spontaneous emission rate}), determines the relaxation time $T_{1}^{\left( \text{vac}\right) }$. For our typical choice of $f_{x} =6.4$~GHz, the corresponding $T_{1}^{\left( \text{vac}\right) }$ is $99$~s. This gives a remarkably long upper bound of the coherence time $T_{2}^{\left( \text{vac}\right) }$ , ref~\cite{traficante1991},
\begin{equation}
	T_{2}^{\left( \text{vac}\right) }\leq 2T_{1}^{\left( \text{vac}\right) }=198~%
	\text{s}. \label{T2v}
\end{equation}

\subsection{Spontaneous emission in a superconducting cavity}

In practice, the electron usually resides in an on-chip superconducting resonator. The resonator alters the spontaneous emission rate due to the modified electromagnetic environment, known as the Purcell effect~\cite{baranov2017}. For a simple two-level charge state coupled to a single-mode resonator with the resonance frequency $f_{\text{r}}=\omega _{\text{r}}/2\pi $, in the dispersive regime (\ie, the coupling strength $g\ll \left\vert \omega _{x}-\omega _{\text{r}}\right\vert$), the modified spontaneous emission rate can be written as
\begin{equation}
	\mathit{\Gamma} ^{\left( \text{cav}\right) }=\frac{g^{2}\kappa }{\left( \omega
		_{x}-\omega _{\text{r}}\right) ^{2}}\text{,} \label{gamma1v}
\end{equation}
Here, $\kappa$ is the resonator linewidth (leakage rate). For the typical experimental settings, with $g/2\pi \sim 5$~MHz\cite{zhou2024electron}, $\kappa/2\pi \sim 0.5$~MHz  and $\left| f_{x}-f_{\text{r}}\right| \sim 500$~MHz, the relaxation time of the electron in the cavity is shortened to $T_{1}^{\left( \text{cav}\right) }=32$~ms. As a result, the coherence time is limited by an upper bound of $T_{2}^{\left( \text{cav}\right) }\leq 64$~ms, which is much shorter than the upper bound in vacuum but still longer than the coherence times limited by other mechanisms below. In addition, the Purcell limited radiative decay can be suppressed by introducing Purcell filter~\cite{reed2010fast} or other readout techniques. 

\section{Decoherence via phonons}
The electron's lateral motion can couple with the longitudinal (acoustic) phonons inside solid Ne. Consequently, the electron's charge state can undergo relaxation and dephasing via emitting phonons. 

The displacement field $\bm{u}(\bm{r},t)$ of solid Ne due to the longitudinal phonons, in a large space of volume $V$ and undisturbed mass density $\rhoNe=1.444$~g~cm$^{-3}$ (ref.~\cite{o2013merck}), can be written as
\begin{align}
\bm{u}(\bm{r},t) = \sum_{\bm{q}} \frac{\bm{q}}{q} \sqrt{\frac{\hbar}{2V\rhoNe\omega_q}}(a_{\bm{q}} \Ee^{-\Ii\omega_q t} - a^\dagger_{-\bm{q}} \Ee^{+\Ii\omega_q t} ) \Ee^{\Ii\bm{q}\cdot\bm{r}}, \label{Eqn:DisplacementFull}
\end{align}
where $\bm{r}=(x,y,z)$ and $\bm{q}=(q_x,q_y,q_z)$ are the 3D positions and wavevectors, respectively; $a_{\bm{q}}$ and $a^\dagger_{\bm{q}} $ and are the phonon annihilation and creation operators of the plane-wave mode with the wavevector $\bm{q}$; $\omega_q=\cNe q$ is the dispersion relation (in the long wavelength limit) that only depends on the magnitude (not direction) of the wavevector, $q=|\bm{q}|=\sqrt{q_x^2+q_y^2+q_z^2}$, $\cNe=1.133\times 10^5$~cm~s$^{-1}$ is the longitudinal sound velocity~\cite{balzer1971}; $\bm{q}/q$ is the unit vector in the longitudinal wavefront direction. 

The velocity field $\bm{v}(\bm{r},t)$ can be derived as
\begin{equation}
\bm{v}(\bm{r},t) = \sum_{\bm{q}} \frac{\bm{q}}{\Ii q} \sqrt{\frac{\hbar\omega_q}{2V\rhoNe}}(a_{\bm{q}} \Ee^{-\Ii\omega_q t} + a^\dagger_{-\bm{q}} \Ee^{+\Ii\omega_q t} ) \Ee^{\Ii\bm{q}\cdot\bm{r}}, \label{Eqn:VelocityFull}
\end{equation}
and the acoustic velocity potential $\mathit{\Phi}(\bm{r},t)$ is
\begin{equation}
\mathit{\Phi}(\bm{r},t) = \sum_{\bm{q}} \frac{1}{q} \sqrt{\frac{\hbar\omega_q}{2V\rhoNe}}(a_{\bm{q}} \Ee^{-\Ii\omega_q t} + a^\dagger_{-\bm{q}} \Ee^{+\Ii\omega_q t} ) \Ee^{\Ii\bm{q}\cdot\bm{r}}, \label{Eqn:ScalarFull}
\end{equation}
which satisfies
\begin{equation}
\bm{v} = -\nabla \mathit{\Phi}.
\end{equation}
The density variation field $\varrho(\bm{r},t) \equiv \rho(\bm{r},t) - \rhoNe$, where $\rho(\bm{r},t)$ is the actual density field, is
\begin{equation}
\varrho(\bm{r},t) = \sum_{\bm{q}} \frac{1}{\Ii}\sqrt{\frac{\hbar\omega_q\rhoNe}{2V\cNe^2}}(a_{\bm{q}} \Ee^{-\Ii\omega_q t} -a^\dagger_{-\bm{q}} \Ee^{+\Ii\omega_q t}) \Ee^{\Ii\bm{q}\cdot\bm{r}}, \label{Eqn:DensityFull}
\end{equation}
which satisfies the acoustic continuity equation,
\begin{equation}
    \partial_t \varrho(\bm{r},t) + \rhoNe \nabla\cdot\bm{v}(\bm{r},t) = 0.
\end{equation}

In our system, phonons are confined in a semi-infinite space, $z<0$, and the system is translationally invariant in $x-y$ plane. We define the 2D lateral positions $\rpar = (x,y)$ and wavevectors $\qpar = (q_x,q_y)$. The density variation field in the Schr\"odinger picture $(t=0)$ is
\begin{align}
	\varrho(\rpar,z) & = \sum_{\qpar,q_z>0} \frac{1}{\Ii} \sqrt{\frac{\hbar\omega_{\qpar,q_z}\rhoNe}{V\cNe^2}} \Ee^{\Ii\qpar\cdot\rpar} \sin(q_zz) \nonumber\\ 
	& \times (\hat{a}_{\qpar,q_z} - \hat{a}^\dagger_{-\qpar,q_z}), \label{Eqn:DensityVariation}
\end{align}
which only includes the modes when the density variation vanishes at the Ne-vacuum interface, $z=0$, where pressure is zero. The summation in Eq.~(\ref{Eqn:DensityVariation}) only includes $q_{z}>0$ to ensure that the modes are counted correctly. Likewise, the acoustic velocity potential is
\begin{align}
	\mathit{\Phi}(\rpar,z) & = \sum_{\qpar,q_z>0} \sqrt{\frac{\hbar\cNe}{V \rhoNe q}} \Ee^{\Ii\qpar\cdot\rpar} \sin(q_zz) \nonumber \\
	& \times (\hat{a}_{\qpar,q_z} + \hat{a}^\dagger_{-\qpar,q_z}), \label{Eqn:VelocityPotential}
\end{align}
and the displacement field is
\begin{align}
	\bm{u}(\rpar,z) & = \sum_{\qpar,q_z>0} \frac{\bm{q}}{q} \sqrt{\frac{\hbar}{V \rhoNe \omega_{\qpar,q_z}}} \Ee^{\Ii\qpar\cdot\rpar} \cos(q_zz) \nonumber \\
	& \times (\hat{a}_{\qpar,q_z} - \hat{a}^\dagger_{-\qpar,q_z}). \label{Eqn:DisplacementAll}
\end{align}

There are two major electron-phonon coupling mechanisms in this system~\cite{schuster2010, dykman2003}: 1. Phonon-induced displacement of the neon surface, which interacts with the electron via the Pauli repulsion and the varied image potential. 2. Phonon-induced modulation of the neon density and hence the dielectric constant, which interacts with the electron via the varied image potential. 

The total Hamiltonian consists of the electron part $\hat{\mathcal{H}}_{\text{el}}$ as given above, the phonon part $\hat{\mathcal{H}}_{\text{ph}}$, and the electron-phonon coupling part $\hat{\mathcal{H}}_{\text{el-ph}}$, 
\begin{equation}
	\hat{\mathcal{H}} = \hat{\mathcal{H}}_{\text{el}} + \hat{\mathcal{H}}_\text{ph} + \hat{\mathcal{H}}_{\text{el-ph}},
\end{equation}
where
\begin{equation}
	\hat{\mathcal{H}}_{\text{ph}} = \sum_{\qpar,q_z}\hbar\omega_{\qpar,q_z} a^\dagger_{\qpar,q_z} a_{\qpar,q_z},
\end{equation}
and
\begin{equation}
	 \hat{\mathcal{H}}_{\text{el-ph}} = \sum_{\qpar,q_z>0} \mathcal{V}_{\qpar,q_z}(\rpar, z) \Ee^{\Ii\qpar\cdot\rpar} (\hat{a}_{\qpar,q_z}-\hat{a}^\dagger_{-\qpar,q_z}). \label{Eqn:EPhCoupling}
\end{equation}
The Hamiltonian is written in the single electron position representation and the second-quantized phonon fields on the plane-wave bases and ignores the zero-point energy. The operators $\hat{a}_{\bm{q}} = \hat{a}_{\qpar,q_z}$ and $\hat{a}_{\bm{q}}^\dagger = \hat{a}_{\qpar,q_z}^\dagger$ are the same phonon annihilation and creation operators as before except being written explicitly in terms of $\qpar$ and $q_z$. $\mathcal{V}_{\qpar,q_z}(\rpar, z)$ is the coupling strength between an electron at the position $(\rpar,z)$ and a plane-wave phonon of the wavevector $(\qpar,q_z)$. $\omega_{\bm{q}} = \omega_{\qpar,q_z}$ is the same phonon dispersion relation as before, $\omega_{\bm{q}} = \omega_{\qpar,q_z} = \cNe q = \cNe \sqrt{\qmpar^2 +q_z^2}$.

The relaxation rate corresponding to electron scattered by a single phonon can be calculated from the Fermi golden rule, we write the electron-phonon coupling Hamiltonian via the modulation of dielectric constant as $\hat{\mathcal{H}}_{\text{el-ph}}^{(\text{mod})}$ and the displacement of neon surface as   $\hat{\mathcal{H}}_{\text{el-ph}}^{(\text{dis})}$, then  the relaxation rate corresponding to each mechanisms and different thermal phonons can be expressed as
\begin{align}
	\mathit{\Gamma}^{(\text{dis})} &= \frac{2\pi}{\hbar} \sum_{\qpar,q_z>0} \left| \langle n_q| \langle 0_x,0_y, 1_z| \hat{\mathcal{H}}^{(\text{dis})}_{\text{el-ph}} |1_x, 0_y, 1_z \rangle |n^\prime_q \rangle \right|^2 \nonumber \\
    &\times \delta(\sum_q\hbar\omega_q n_q - \sum_q\hbar\omega_q n^\prime_q -\hbar \omega_x ), \nonumber \\ 
    &=\frac{2\pi}{\hbar} \sum_{\qpar,q_z>0} \left| \mathcal{V}^{\text{(dis)}}_{\qpar,q_z}(\rpar,z) \right|_{10}^2 G_{\qpar} \nonumber \\
    & \times\delta(\hbar\omega_{\qpar,q_z} - \hbar\omega_x),\label{GammaDis}\\	
	\mathit{\Gamma}^{(\text{mod})} 
    &=\frac{2\pi}{\hbar} \sum_{\qpar,q_z>0} \left| \mathcal{V}^{\text{(dis)}}_{\qpar,q_z}(\rpar,z) \right|_{10}^2 G_{\qpar}  \nonumber \\
    & \times\delta(\hbar\omega_{\qpar,q_z} - \hbar\omega_x),  \label{GammaMod}
\end{align}
where the subscript ``10" denotes the excited to ground state transition and $n_q=1/[\exp(\hbar\omega_q/(\KB T))-1] \ll 1$ denotes the average thermal occupation number of the phonon state with the wavenumber $q$ at $T=10$~mK. The function $G_{\qpar}$ is
\begin{align}
	G_{\qpar} & = |\langle 0_x, 0_y, 1_z|\Ee^{\Ii\qpar\cdot\rpar}|1_x, 0_y, 1_z\rangle|^2 \nonumber \\
	& = \frac{1}{2}\left(q_x a_x\right)^2\Ee^{-(q_x^2a_x^2+q_y^2a_y^2)/2},
\end{align}
where $a_x = \sqrt{\hbar/m_e \omega_x}$, $a_y = \sqrt{\hbar/m_e \omega_y}$ are the characteristic trapping widths in $x$ and $y$, respectively.

The phonon-induced dephasing ($\Gamma_\phi$) of electrons was estimated following the treatment used in ref~\cite{dykman2003} based on the ensemble-averaged mean-square phase increment:
\begin{align}
    \overline {\langle [\phi_{10}(t)-\phi_{10}(0)]^2 \rangle }&= \frac{1}{\hbar^2}\int_0^t dt^{\prime}\int_0^t dt^{\prime\prime}\overline{\langle E_{10}(t^{\prime})E_{10}(t^{\prime\prime})  \rangle} \\ \nonumber
    &=\int_0^t dt^{\prime}\int_0^t dt^{\prime\prime} \Gamma_\phi \delta (t^\prime -t^{\prime\prime}) = \Gamma_\phi t,
\end{align}
where $\overline {\langle \cdot \rangle}$ is the average over the thermal states of phonons and 
\begin{align}
    E_{10}(t) &=  \langle 1_x, 0_y, 1_z| \hat{\mathcal{H}}_{\text{el-ph}}(t) |1_x, 0_y, 1_z \rangle \nonumber\\
    &-\langle 0_x, 0_y, 1_z| \hat{\mathcal{H}}_{\text{el-ph}}(t) |0_x, 0_y, 1_z \rangle, 
    \label{eq:depasing}
\end{align}
is the energy difference between two electronic states. The energy correlator can be approximated by a $\delta$-function which gives the depahsing rate $\Gamma_\phi$. The average value of the energy difference Eq.~(\ref{eq:depasing}) is given by
\begin{align}
   &  \overline{\langle E_{10}(t^{\prime})E_{10}(t^{\prime\prime})  \rangle} \\ \nonumber
    &=\sum_{\qpar,q_z>0}  \left[ (\mathcal{V}_{\qpar,q_z>0}\Ee^{\Ii\qpar\cdot\rpar})_{11} - (\mathcal{V}_{\qpar,q_z>0}\Ee^{\Ii\qpar\cdot\rpar})_{00} \right]^2 \\ \nonumber
        &\times [(n_q+1)e^{-i \omega_q(t^{\prime}-t^{\prime\prime})} - n_qe^{i \omega_q (t^{\prime}-t^{\prime\prime})}],
\end{align}
and the integral over time can be found as
\begin{align}
    \int_0^t dt^{\prime}\int_0^t dt^{\prime\prime} e^{-i\omega_q(t^{\prime}-t^{\prime\prime})} = \frac{4\sin^2(\omega_q t/2)}{\omega^2_q} \\ \nonumber
    =\lim_{t \rightarrow \infty} 2\pi t \delta(\omega_q),
\end{align}
with $\delta(\omega_q)=c^{-1}_{\text{Ne}}\delta(q)$ ,which result in 
\begin{align}
    \mathit{\Gamma}_\phi = \frac{2\pi}{\hbar^2 c_{\text{Ne}}}\sum_{\qpar,q_z>0}  [ &(\mathcal{V}_{\qpar,q_z>0}\Ee^{\Ii\qpar\cdot\rpar})_{11} - \\ \nonumber
    &-(\mathcal{V}_{\qpar,q_z>0}\Ee^{\Ii\qpar\cdot\rpar})_{00} ]^2 \delta(q).
\end{align}
For one-phonon processes, the summation effectively vanishes due to the presence of $\delta(q)$, indicating that pure dephasing is negligible. Consequently, the decoherence time is governed by the relaxation process. In the following section, we focus on calculating the relaxation times for two distinct interaction mechanisms.

\subsection{Phonon-induced displacement of neon surface}

The electron-phonon coupling strength $\mathcal{V}_{\qpar,q_z}(\rpar,z)$ in Eq.~(\ref{Eqn:EPhCoupling}) through the mechanism of phonon-induced displacement of neon surface is given by
\begin{align}
	\mathcal{V}^{\text{(dis)}}_{\qpar,q_z}(\rpar,z)
	& = \frac{q_z}{q}\sqrt{\frac{\hbar}{V \rhoNe \cNe q}} \left\{ -\frac{\Ii}{m_e}(\qpar\cdot\oppar)\hat{p}_z \right. \nonumber\\ 
	& \left. - \frac{\Ii\hbar}{2m_e} \qmpar^2\hat{p}_z + e\mathcal{E}_z +  \mathit{\Lambda} \qmpar^2U_{\text{p}}(\qmpar z) \right\},
\end{align}
Here $\oppar=-i\hbar \partial _{\bm{r}}$ is the in-plane electron momentum operator, $\hat{p}_{z}=-i\hbar \partial _{z}$ is the vertical electron momentum operator, $q_z/q$ restrict ourselves to the vertical displacement of Ne surface.

The $U_{\text{p}}(\eta)$ function is associated with the change of polarization potential due to surface curvature~\cite{saitoh1977,monarkha1982},
\begin{equation}
	U_{\text{p}}(\eta) = \frac{1}{\eta^2} \left[ 1 - \eta K_1(\eta) \right].
\end{equation}
Here  $K_{1}(\eta)$ is the modified Bessel function of the first kind. The first two terms describe the kinematic interaction arising from the electron's wavefunction being set to zero on a non-flat hard-wall surface. The first two terms have nonzero contribution only if we are considering out-of-plane excitation. The pressing electric field term in practical experiments is much smaller than the polarization term, so only the polarization term needs to be included in our calculation. The relaxation rate can be simplified as
\begin{align}
	\mathit{\Gamma}^{\text{(dis)}} & = \frac{2\pi}{\hbar} \frac{\hbar\cNe}{V \rhoNe} \sum_{\qpar,q_z>0} \frac{q_z^2}{q^2} \frac{1}{q \cNe^2} G_{\qpar} U_{\qpar,q_z} \nonumber \\ 
	& \times\delta(\hbar\omega_{\qpar,q_z} - \hbar\omega_x),
\end{align}
where
\begin{align}
	U_{\qpar,q_z} & = \left|\mathit{\Lambda} \qmpar^2 \langle 1_z | U_\text{p}(\qmpar z) | 1_z \rangle \right|^2 \nonumber \\
	& = \left|\mathit{\Lambda} \qmpar^2 \int\Dd z~ |\psi_1(z)|^2 \frac{1}{(\qmpar z)^2} \left[1-\qmpar zK_1(\qmpar z)\right] \right|^2 \nonumber \\
	& \approx -\frac{1}{2}\mathit{\Lambda}^2 \qmpar^4 \ln (\qmpar\rB),
\end{align}
with the approximation $\qmpar z\rightarrow0$ for our problem.

Finally, the decay rate can be written as
\begin{align}
\mathit{\Gamma}^{(\text{dis})} & = \frac{R ^{2}\rB^{2}\omega_x^6}{8\pi m_e \rhoNe \cNe^{9}} \int_{0}^{1} \Dd\gamma~ \gamma^2(1-\gamma^2)^3 \nonumber \\ 
& \times \exp\left[-\frac{\hbar\omega_x}{2m_e\cNe^2} (1-\gamma^2)\right] \nonumber \\
& \times \left[\ln \left( \frac{\omega_x}{\cNe}\rB\sqrt{1-\gamma^2} \right) \right]^2 . \label{gamma1_pol}
\end{align}%
For simplicity, we have assumed $\omega_{x}=\omega_{y}$. The decay rate $\mathit{\Gamma}^{\left(\text{dis}\right)}$ is solely determined by the confinement frequency $\omega_{x}$. 
Numerical integral shows $\mathit{\Gamma}^{(\text{dis})} = 49.4$~s$^{-1}$ for the transition frequency at $f_x= 6.4$~GHz. This leads to the corresponding relaxation time,
\begin{equation}
	T_{1}^{\left( \text{dis}\right) } (6.4~\text{GHz}) =1/\mathit{\Gamma} ^{\left( \text{dis}\right) }=20.2%
	\text{ ms},  \label{T_1s}
\end{equation}
and coherence time,
\begin{equation}
	T_{2}^{\left( \text{dis}\right) } (6.4~\text{GHz}) =\left( \frac{\mathit{\Gamma} ^{\left( \text{dis}%
			\right) }}{2}+\Gamma _{\phi }\right) ^{-1}=40.4\text{ ms},  \label{T_2s}
\end{equation}
both of which are much longer than the reported observations.

\begin{table}[tbp]
	\caption{Calculated relaxation time $T_{1}^{\left( \text{dis}\right) }$ and coherence time $T_{2}^{\left( \text{dis}\right) }$ of an electron in-plane motional state with varied transition frequency $f_{x}$ for phonon-induced surface displacement.}
	\label{tab:table1}%
	\begin{ruledtabular}
		\begin{tabular}{llll}
			\textrm{$f_x$ (GHz)}&
			\textrm{$T_{1}^{\left( \text{dis}\right) }$ (s)}&
			\textrm{$T_{2}^{\left( \text{dis}\right) }$ (s)}\\
			\colrule
			1 & $183.1$ & $366.2$ \\
			2 & $4.79$ & $9.58$ \\
			3 & $0.63$ & $1.26$ \\
			4 & $0.16$ & $0.32$ \\
			5 & $0.06$ & $0.12$ \\
			6 & $0.026$  & $5.31\times 10^{-2}$\\
			7 & $0.014$ & $2.81\times 10^{-2}$ \\
			8 & $8.3\times 10^{-3}$ & $1.66\times 10^{-2}$ \\
			9 & $5.3\times 10^{-3}$  & $1.06\times 10^{-2}$\\
			10& $3.6\times 10^{-3}$ & $7.29\times 10^{-3}$ \\
			
		\end{tabular}
	\end{ruledtabular}
\end{table}
In Table~\ref{tab:table1} and Fig.2, we present the variations of $T_{1}^{\left(\text{dis}\right)}$ and $T_{2}^{\left(\text{dis}\right)}$ with different transition frequency $f_{x}$. In most cases, $T_{1}^{\left(\text{dis}\right)}$ and $T_{2}^{\left(\text{dis}\right)}$ exceed 1~ms, which surpass the $T_{1}$ and $T_{2}$ values for most semiconductor and superconducting charge qubits.

\subsection{Phonon-induced modulation of dielectric constant }
Another decoherence mechanism comes from the phonon-induced modulation of dielectric constant $\hat{\mathcal{H}}_{\text{el-ph}}^{(\text{mod})}$. The dielectric constant variation $\delta \epsilon$ and the density variation $\varrho$ is related by 
\begin{equation}
 \delta \epsilon = (\epsNe-1) \frac{\varrho}{\rhoNe}.
\end{equation}

The electron-phonon coupling strength $\mathcal{V}_{\qpar,q_z}(\rpar,z)$ in Eq.~(\ref{Eqn:EPhCoupling}) through the mechanism of phonon-induced modulation of dielectric constant is given by
\begin{align}
	\mathcal{V}^{\text{(mod)}}_{\qpar,q_z}(z) & = -\Ii \mathit{\Lambda} \qmpar \sqrt{\frac{\hbar\omega_{\qpar,q_z}}{V \rhoNe \cNe^2}} \int_{-\infty}^{0} \Dd z'~ \frac{\sin(q_z z')}{z-z'} \nonumber \\ & \times K_1(\qmpar|z-z'|), \label{Eqn:EPhCouplingFourier}
\end{align}
which is independent of $\rpar$ and contains the modified Bessel function of the first kind, $K_{1}(\eta)$. The relaxation rate can be calculated with Eq.~(\ref{GammaMod}),
\begin{align}
	\mathit{\Gamma}^{\text{(mod)}} 
	& = \frac{2\pi}{\hbar} \frac{\mathit{\Lambda}^2}{V \rhoNe\cNe^2} \sum_{\bm{q},q_z>0} q^2\hbar\omega_{\qpar,q_z} G_{\qpar} F_{\qpar,q_z} \nonumber \\ 
	& \times\delta(\hbar\omega_{\qpar,q_z} - \hbar\omega_x) .
\end{align}
 The function $F_{\qpar,q_z}$ is
\begin{align}
	F_{\qpar,q_z} & = \left| \langle 1_z | \int_{-\infty}^{0} \Dd z'~ \frac{\sin(q_z z')}{z-z'} K_1(\qmpar|z-z'|) | 1_z \rangle \right|^2 \nonumber\\
& = \left| \int_0^{\infty}\Dd z~ |\psi_1(z)|^2 \right. \nonumber\\
	& \left. \times\int_{-\infty}^{0} \Dd z'~ \frac{\sin(q_z z')}{z-z'} K_1(\qmpar|z-z'|) \right|^2, 
\end{align}
where $\psi_1(z)$ is the normalized electron's ground-state wavefunction in $z$ given by Eq.~(\ref{Eqn:RydbergGround}).
To calculate the integrals in $F_{\qpar,q_z}$, note that $z>0$ for the electron and $z'<0$ for the image charges, therefore $z-z'= |z-z'| >0$ always holds. 

Finally, the decay rate can be written as
\begin{align}
	\mathit{\Gamma}^{\text{(mod)}} & = \frac{8R^2\omega_x^4}{\pi m_e\rhoNe\cNe^7} \int_{0}^{1} \Dd \gamma ~(1-\gamma^2) \nonumber \\ 
	& \times \exp\left[-\frac{\hbar\omega_x}{2m_e\cNe^2} (1-\gamma^2)\right] \nonumber \\
	& \times \left| \int_0^{\infty}\Dd s \int_{0}^{\infty} \Dd s'~ \frac{s^2}{(s+s')^2} \sin\left(\frac{\omega_x}{\cNe}\rB u s'\right)\Ee^{-2s} \right|^2, ~\label{Eqn:GammaMod}
\end{align}
where $\gamma\equiv \cos\vartheta$ with $\vartheta$ being the polar angle of phonon wavevectors, $s\equiv z/\rB>0$, and $s'\equiv -z'/\rB>0$. They are all positive dimensionless variables and Eq.~(\ref{Eqn:GammaMod}) can be numerically integrated without further approximations.

\begin{figure}[htb]
	\begin{center}
		\includegraphics[scale=0.7]{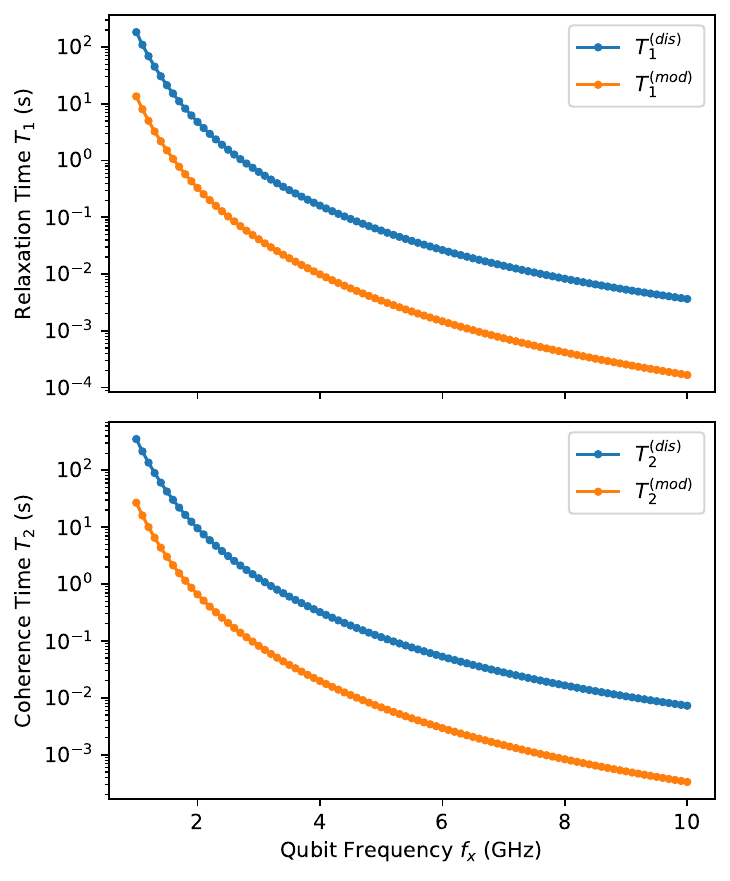}
	\end{center}
	\caption{Calculated relaxation time $T_1$ (top) and coherence time $T_2$ (bottom) for the two electron-phonon coupling mechanisms: phonon-induced surface displacement and phonon-induced dielectric-constant modulation. }
	\label{Fig:2-T1}
\end{figure}

For a qubit transition frequency of $f_x= \omega_x/2\pi = 6.4$~GHz, as commonly used in our experiments, ignoring the practical need of anharmonicity, the numerical integration gives a relaxation rate of $\mathit{\Gamma}^{\text{(mod)}} = 908$~s$^{-1}$. Consequently, the relaxation time $T_1$ is
\begin{equation}
T_{1}^{\text{(mod)}} (6.4~\text{GHz}) = \frac{1}{\mathit{\Gamma}^{\text{(mod)}}} = 1.1~\text{ms}. \label{T1}
\end{equation}
and the coherence time $T_2$ is relaxation-limited~\cite{chen2022},
\begin{equation}
	T_{2}^{\text{(mod)}} (6.4~\text{GHz}) =\left( \frac{\mathit{\Gamma}^{\text{(mod)}}}{2}+\mathit{\Gamma} _{\phi }\right) ^{-1}=2.2~\text{ms}.  \label{T2}
\end{equation}
These calculated numbers are at least one order of magnitude better than our experimentally observed $\sim$0.1~ms. In Table~\ref{tab:table2} and Fig.~\ref{Fig:2-T1}, we present how $T_{1}$ and $T_{2}$ vary with the qubit transition frequency $f_x$. In all cases, they are longer than 0.1~ms, and longer than those of most semiconductor charge qubits.
\begin{table}[tbp]
	\caption{Calculated relaxation time $T_1^{\text{(mod)}}$ and	coherence time $T_2^{\text{(mod)}}$ of an electron in-plane motional state with varied transition frequency $f_x$ for phonon-induced dielectric-constant modulation. }
	\label{tab:table2}%
	\begin{ruledtabular}
		\begin{tabular}{llll}
			$f_x$ (GHz) &
			$T_1^{\text{(mod)}}$ (s) &
			$T_2^{\text{(mod)}} (s)$ \\
			\colrule
			1 & $13.49$ & $27.0$ \\
			2 & $0.33$ & $0.66$ \\
			3 & $0.041$ & $0.082$ \\
			4 & $0.010$ & $1.98\times 10^{-2}$ \\
			5 & $3.4\times 10^{-3}$ & $6.81\times 10^{-3}$ \\
			6 & $1.5\times 10^{-3}$  & $2.94\times 10^{-3}$\\
			7 & $7.4\times 10^{-4}$ & $1.48\times 10^{-3}$ \\
			8 & $4.2\times 10^{-4}$ & $8.34\times 10^{-4}$ \\
			9 & $2.6\times 10^{-4}$  & $5.11\times 10^{-4}$\\
			10 & $1.7\times 10^{-4}$ & $3.33\times 10^{-4}$ \\		
		\end{tabular}
	\end{ruledtabular}
\end{table}

So far, we have assumed that the solid Ne occupies the entire lower half space $z<0$. This is a valid approximation when the Ne film is thicker than $\sim$10~nm. However, if the Ne film is only a few nanometers thick, the image charge will be dominated by the underlying dielectric substrate, such as a silicon (Si) or sapphire. Then the electron decoherence is primarily induced by the phonons in Si or sapphire. At low temperatures, the longitudinal sound velocity in silicon and sapphire are $c_{\text{Si}} = 8480$ m/s (ref.~\cite{mcskimin1964}) and $c_{\text{Sa}} = 11350$ m/s (ref.~\cite{sarbei1979}), respectively. Consequently, the typical electron's in-plane wavenumber $\left(m\omega_x/\hbar\right)^{1/2} = 1.9\times 10^7$ m$^{-1}$ is much larger than the wavenumbers of the resonant phonons $\omega_x/c_{\text{Si}} = 4.7\times10^6$ m$^{-1}$ and $ \omega_x/c_{\text{Sa}} = 3.5\times 10^6$ m$^{-1}$ ($f_x=6.4$~GHz). 
This mismatch makes the relaxation rate through these channels essentially 0, following the above calculation. Physically it originates from an inability to simultaneously satisfy energy and momentum conservation during the scattering process. Therefore, the single phonon scattering from the substrate is exponentially suppressed due to the mismatch between the size of the electron wave function and the phonon wavelength at the same energy. In experiments, the thickness of the Ne film typically lies between the two aforementioned extremes. As a result, the actual electron charge coherence time may exceed our conservative estimations above.

\section{Conclusion}

In this theoretical study, we have investigated the charge decoherence of a single electron on a solid neon surface interacting with the acoustic phonons in the solid neon. We find in the transition frequency range from 1~GHz to 10~GHz, for the phonon-induced displacement of the Ne surface, the coherence time $T_{2}^{\left( \text{dis}\right) }$ range from approximately 366~s to 7~ms, and for the phonon-induced modulations of Ne relative permittivity, the coherence tim $T_{2}^{\left( \text{mod}\right)} $ ranges from 27~s to 0.3~ms. These results highlight the promise of single-electron motional (charge) qubits on solid Ne as a long-coherence platform, calling for efforts to improve the experimentally achieved coherence performance.

\begin{acknowledgments}

D. J., X. L., and Q. C. acknowledge support from the Argonne National Laboratory Directed Research and Development (LDRD) program. D. J., X. L., and S. Z. acknowledge support from the Air Force Office of Scientific Research (AFOSR) under Award No. FA9550-23-1-0636. D. J. acknowledges support from the Department of Energy (DOE) under Award No. DE-SC0025542 and the National Science Foundation (NSF) under Award No. OSI-2426768. Work performed at the Center for Nanoscale Materials, a U.S. Department of Energy Office of Science User Facility, was supported by the U.S. DOE, Office of Basic Energy Sciences, under Contract No. DEAC02-06CH11357. The authors thank Mark I. Dykman for helpful discussions.

\end{acknowledgments}



\nocite{*}

\bibliography{EMCRef}

\providecommand{\noopsort}[1]{}\providecommand{\singleletter}[1]{#1}%
\begin{thebibliography}{39}%
\makeatletter
\providecommand \@ifxundefined [1]{%
 \@ifx{#1\undefined}
}%
\providecommand \@ifnum [1]{%
 \ifnum #1\expandafter \@firstoftwo
 \else \expandafter \@secondoftwo
 \fi
}%
\providecommand \@ifx [1]{%
 \ifx #1\expandafter \@firstoftwo
 \else \expandafter \@secondoftwo
 \fi
}%
\providecommand \natexlab [1]{#1}%
\providecommand \enquote  [1]{``#1''}%
\providecommand \bibnamefont  [1]{#1}%
\providecommand \bibfnamefont [1]{#1}%
\providecommand \citenamefont [1]{#1}%
\providecommand \href@noop [0]{\@secondoftwo}%
\providecommand \href [0]{\begingroup \@sanitize@url \@href}%
\providecommand \@href[1]{\@@startlink{#1}\@@href}%
\providecommand \@@href[1]{\endgroup#1\@@endlink}%
\providecommand \@sanitize@url [0]{\catcode `\\12\catcode `\$12\catcode `\&12\catcode `\#12\catcode `\^12\catcode `\_12\catcode `\%12\relax}%
\providecommand \@@startlink[1]{}%
\providecommand \@@endlink[0]{}%
\providecommand \url  [0]{\begingroup\@sanitize@url \@url }%
\providecommand \@url [1]{\endgroup\@href {#1}{\urlprefix }}%
\providecommand \urlprefix  [0]{URL }%
\providecommand \Eprint [0]{\href }%
\providecommand \doibase [0]{https://doi.org/}%
\providecommand \selectlanguage [0]{\@gobble}%
\providecommand \bibinfo  [0]{\@secondoftwo}%
\providecommand \bibfield  [0]{\@secondoftwo}%
\providecommand \translation [1]{[#1]}%
\providecommand \BibitemOpen [0]{}%
\providecommand \bibitemStop [0]{}%
\providecommand \bibitemNoStop [0]{.\EOS\space}%
\providecommand \EOS [0]{\spacefactor3000\relax}%
\providecommand \BibitemShut  [1]{\csname bibitem#1\endcsname}%
\let\auto@bib@innerbib\@empty
\bibitem [{\citenamefont {Chatterjee}\ \emph {et~al.}(2021)\citenamefont {Chatterjee}, \citenamefont {Stevenson}, \citenamefont {Franceschi}, \citenamefont {A.~Morello},\ and\ \citenamefont {Kuemmeth}}]{chatterjee2021}%
  \BibitemOpen
  \bibfield  {author} {\bibinfo {author} {\bibfnamefont {A.}~\bibnamefont {Chatterjee}}, \bibinfo {author} {\bibfnamefont {P.}~\bibnamefont {Stevenson}}, \bibinfo {author} {\bibfnamefont {S.~D.}\ \bibnamefont {Franceschi}}, \bibinfo {author} {\bibfnamefont {N.~d.~L.}\ \bibnamefont {A.~Morello}},\ and\ \bibinfo {author} {\bibfnamefont {F.}~\bibnamefont {Kuemmeth}},\ }\href@noop {} {\bibfield  {journal} {\bibinfo  {journal} {Nat. Rev. Phys.}\ }\textbf {\bibinfo {volume} {3}},\ \bibinfo {pages} {157} (\bibinfo {year} {2021})}\BibitemShut {NoStop}%
\bibitem [{\citenamefont {Siddiqi}(2021)}]{siddiqi2021engineering}%
  \BibitemOpen
  \bibfield  {author} {\bibinfo {author} {\bibfnamefont {I.}~\bibnamefont {Siddiqi}},\ }\bibfield  {title} {\enquote {\bibinfo {title} {Engineering high-coherence superconducting qubits},}\ }\href@noop {} {\bibfield  {journal} {\bibinfo  {journal} {Nature Reviews Materials}\ }\textbf {\bibinfo {volume} {6}},\ \bibinfo {pages} {875--891} (\bibinfo {year} {2021})}\BibitemShut {NoStop}%
\bibitem [{\citenamefont {Veldhorst}\ \emph {et~al.}(2014)\citenamefont {Veldhorst}, \citenamefont {Hwang}, \citenamefont {Yang}, \citenamefont {Leenstra}, \citenamefont {de~Ronde}, \citenamefont {Dehollain}, \citenamefont {Muhonen}, \citenamefont {Hudson}, \citenamefont {Itoh}, \citenamefont {Morello},\ and\ \citenamefont {Dzurak}}]{veldhorst2014}%
  \BibitemOpen
  \bibfield  {author} {\bibinfo {author} {\bibfnamefont {M.}~\bibnamefont {Veldhorst}}, \bibinfo {author} {\bibfnamefont {J.~C.~C.}\ \bibnamefont {Hwang}}, \bibinfo {author} {\bibfnamefont {C.~H.}\ \bibnamefont {Yang}}, \bibinfo {author} {\bibfnamefont {A.~W.}\ \bibnamefont {Leenstra}}, \bibinfo {author} {\bibfnamefont {B.}~\bibnamefont {de~Ronde}}, \bibinfo {author} {\bibfnamefont {J.~P.}\ \bibnamefont {Dehollain}}, \bibinfo {author} {\bibfnamefont {J.~T.}\ \bibnamefont {Muhonen}}, \bibinfo {author} {\bibfnamefont {F.~E.}\ \bibnamefont {Hudson}}, \bibinfo {author} {\bibfnamefont {K.~M.}\ \bibnamefont {Itoh}}, \bibinfo {author} {\bibfnamefont {A.}~\bibnamefont {Morello}},\ and\ \bibinfo {author} {\bibfnamefont {A.~S.}\ \bibnamefont {Dzurak}},\ }\bibfield  {title} {\enquote {\bibinfo {title} {An addressable quantum dot qubit with fault-tolerant control-fidelity},}\ }\href@noop {} {\bibfield  {journal} {\bibinfo  {journal} {Nat. Nanotechnol.}\ }\textbf {\bibinfo {volume} {9}},\ \bibinfo {pages} {981} (\bibinfo
  {year} {2014})}\BibitemShut {NoStop}%
\bibitem [{\citenamefont {K.Takeda}\ \emph {et~al.}(2016)\citenamefont {K.Takeda}, \citenamefont {Kamioka}, \citenamefont {Otsuka}, \citenamefont {J.Yoneda}, \citenamefont {Nakajima}, \citenamefont {Delbecq}, \citenamefont {Amaha}, \citenamefont {Allison}, \citenamefont {Kodera}, \citenamefont {Oda},\ and\ \citenamefont {S.Tarucha}}]{takeda2016}%
  \BibitemOpen
  \bibfield  {author} {\bibinfo {author} {\bibnamefont {K.Takeda}}, \bibinfo {author} {\bibfnamefont {J.}~\bibnamefont {Kamioka}}, \bibinfo {author} {\bibfnamefont {T.}~\bibnamefont {Otsuka}}, \bibinfo {author} {\bibnamefont {J.Yoneda}}, \bibinfo {author} {\bibfnamefont {T.}~\bibnamefont {Nakajima}}, \bibinfo {author} {\bibfnamefont {M.~R.}\ \bibnamefont {Delbecq}}, \bibinfo {author} {\bibfnamefont {S.}~\bibnamefont {Amaha}}, \bibinfo {author} {\bibfnamefont {G.}~\bibnamefont {Allison}}, \bibinfo {author} {\bibfnamefont {T.}~\bibnamefont {Kodera}}, \bibinfo {author} {\bibfnamefont {S.}~\bibnamefont {Oda}},\ and\ \bibinfo {author} {\bibnamefont {S.Tarucha}},\ }\href@noop {} {\bibfield  {journal} {\bibinfo  {journal} {Sci. Adv.}\ }\textbf {\bibinfo {volume} {2}},\ \bibinfo {pages} {e1600694} (\bibinfo {year} {2016})}\BibitemShut {NoStop}%
\bibitem [{\citenamefont {Osman}\ \emph {et~al.}(2021)\citenamefont {Osman}, \citenamefont {Simon}, \citenamefont {Bengtsson}, \citenamefont {Kosen}, \citenamefont {Krantz}, \citenamefont {Lozano}, \citenamefont {Scigliuzzo}, \citenamefont {Delsing}, \citenamefont {Bylander},\ and\ \citenamefont {Roudsari}}]{osman2021}%
  \BibitemOpen
  \bibfield  {author} {\bibinfo {author} {\bibfnamefont {A.}~\bibnamefont {Osman}}, \bibinfo {author} {\bibfnamefont {J.}~\bibnamefont {Simon}}, \bibinfo {author} {\bibfnamefont {A.}~\bibnamefont {Bengtsson}}, \bibinfo {author} {\bibfnamefont {S.}~\bibnamefont {Kosen}}, \bibinfo {author} {\bibfnamefont {P.}~\bibnamefont {Krantz}}, \bibinfo {author} {\bibfnamefont {D.}~\bibnamefont {Lozano}}, \bibinfo {author} {\bibfnamefont {M.}~\bibnamefont {Scigliuzzo}}, \bibinfo {author} {\bibfnamefont {P.}~\bibnamefont {Delsing}}, \bibinfo {author} {\bibfnamefont {J.}~\bibnamefont {Bylander}},\ and\ \bibinfo {author} {\bibfnamefont {A.~F.}\ \bibnamefont {Roudsari}},\ }\href@noop {} {\bibfield  {journal} {\bibinfo  {journal} {Appl. Phys. Lett.}\ }\textbf {\bibinfo {volume} {118}},\ \bibinfo {pages} {064002} (\bibinfo {year} {2021})}\BibitemShut {NoStop}%
\bibitem [{\citenamefont {Pashkin}\ \emph {et~al.}(2009)\citenamefont {Pashkin}, \citenamefont {Astafiev}, \citenamefont {T.Yamamoto}, \citenamefont {Nakamura},\ and\ \citenamefont {Tsai}}]{pashkin2009}%
  \BibitemOpen
  \bibfield  {author} {\bibinfo {author} {\bibfnamefont {Y.~A.}\ \bibnamefont {Pashkin}}, \bibinfo {author} {\bibfnamefont {O.}~\bibnamefont {Astafiev}}, \bibinfo {author} {\bibnamefont {T.Yamamoto}}, \bibinfo {author} {\bibfnamefont {Y.}~\bibnamefont {Nakamura}},\ and\ \bibinfo {author} {\bibfnamefont {J.~S.}\ \bibnamefont {Tsai}},\ }\href@noop {} {\bibfield  {journal} {\bibinfo  {journal} {Quantum Inf. Process.}\ }\textbf {\bibinfo {volume} {8}},\ \bibinfo {pages} {55} (\bibinfo {year} {2009})}\BibitemShut {NoStop}%
\bibitem [{\citenamefont {Samkharadze}\ \emph {et~al.}(2016)\citenamefont {Samkharadze}, \citenamefont {Bruno}, \citenamefont {Scarlino}, \citenamefont {Zheng}, \citenamefont {DiVincenzo}, \citenamefont {DiCarlo},\ and\ \citenamefont {L.Vandersypen}}]{diVincenzo2016}%
  \BibitemOpen
  \bibfield  {author} {\bibinfo {author} {\bibfnamefont {N.}~\bibnamefont {Samkharadze}}, \bibinfo {author} {\bibfnamefont {A.}~\bibnamefont {Bruno}}, \bibinfo {author} {\bibfnamefont {P.}~\bibnamefont {Scarlino}}, \bibinfo {author} {\bibfnamefont {G.}~\bibnamefont {Zheng}}, \bibinfo {author} {\bibfnamefont {D.}~\bibnamefont {DiVincenzo}}, \bibinfo {author} {\bibfnamefont {L.}~\bibnamefont {DiCarlo}},\ and\ \bibinfo {author} {\bibnamefont {L.Vandersypen}},\ }\href@noop {} {\bibfield  {journal} {\bibinfo  {journal} {Phys. Rev. Appl.}\ }\textbf {\bibinfo {volume} {5}},\ \bibinfo {pages} {044004} (\bibinfo {year} {2016})}\BibitemShut {NoStop}%
\bibitem [{\citenamefont {Kroll}\ \emph {et~al.}(2019)\citenamefont {Kroll}, \citenamefont {Borsoi}, \citenamefont {Enden}, \citenamefont {Uilhoorn}, \citenamefont {Jong}, \citenamefont {Quintero-P\'erez}, \citenamefont {D.VanWoerkom}, \citenamefont {Bruno}, \citenamefont {Plissard}, \citenamefont {Car}, \citenamefont {Bakkers}, \citenamefont {Cassidy},\ and\ \citenamefont {Kouwenhoven}}]{Kroll2019}%
  \BibitemOpen
  \bibfield  {author} {\bibinfo {author} {\bibfnamefont {J.}~\bibnamefont {Kroll}}, \bibinfo {author} {\bibfnamefont {F.}~\bibnamefont {Borsoi}}, \bibinfo {author} {\bibfnamefont {K.~D.}\ \bibnamefont {Enden}}, \bibinfo {author} {\bibfnamefont {W.}~\bibnamefont {Uilhoorn}}, \bibinfo {author} {\bibfnamefont {D.~D.}\ \bibnamefont {Jong}}, \bibinfo {author} {\bibfnamefont {M.}~\bibnamefont {Quintero-P\'erez}}, \bibinfo {author} {\bibnamefont {D.VanWoerkom}}, \bibinfo {author} {\bibfnamefont {A.}~\bibnamefont {Bruno}}, \bibinfo {author} {\bibfnamefont {S.}~\bibnamefont {Plissard}}, \bibinfo {author} {\bibfnamefont {D.}~\bibnamefont {Car}}, \bibinfo {author} {\bibfnamefont {E.}~\bibnamefont {Bakkers}}, \bibinfo {author} {\bibfnamefont {M.}~\bibnamefont {Cassidy}},\ and\ \bibinfo {author} {\bibfnamefont {L.}~\bibnamefont {Kouwenhoven}},\ }\href@noop {} {\bibfield  {journal} {\bibinfo  {journal} {Phys. Rev. Appl.}\ }\textbf {\bibinfo {volume} {11}},\ \bibinfo {pages} {064053} (\bibinfo {year} {2019})}\BibitemShut
  {NoStop}%
\bibitem [{\citenamefont {Stano}\ and\ \citenamefont {Loss}(2022)}]{loss2022}%
  \BibitemOpen
  \bibfield  {author} {\bibinfo {author} {\bibfnamefont {P.}~\bibnamefont {Stano}}\ and\ \bibinfo {author} {\bibfnamefont {D.}~\bibnamefont {Loss}},\ }\href@noop {} {\bibfield  {journal} {\bibinfo  {journal} {Nat. Rev. Phys.}\ }\textbf {\bibinfo {volume} {4}},\ \bibinfo {pages} {672} (\bibinfo {year} {2022})}\BibitemShut {NoStop}%
\bibitem [{\citenamefont {Blais}, \citenamefont {Grimsmo},\ and\ \citenamefont {A.Wallraff}(2021)}]{blais2021}%
  \BibitemOpen
  \bibfield  {author} {\bibinfo {author} {\bibfnamefont {A.}~\bibnamefont {Blais}}, \bibinfo {author} {\bibfnamefont {A.~L.}\ \bibnamefont {Grimsmo}},\ and\ \bibinfo {author} {\bibnamefont {A.Wallraff}},\ }\href@noop {} {\bibfield  {journal} {\bibinfo  {journal} {Rev. Mod. Phys.}\ }\textbf {\bibinfo {volume} {93}},\ \bibinfo {pages} {025005} (\bibinfo {year} {2021})}\BibitemShut {NoStop}%
\bibitem [{\citenamefont {D'Anjou}\ and\ \citenamefont {Burkard}(2019)}]{burkard2019}%
  \BibitemOpen
  \bibfield  {author} {\bibinfo {author} {\bibfnamefont {B.}~\bibnamefont {D'Anjou}}\ and\ \bibinfo {author} {\bibfnamefont {G.}~\bibnamefont {Burkard}},\ }\href@noop {} {\bibfield  {journal} {\bibinfo  {journal} {Phys. Rev. B}\ }\textbf {\bibinfo {volume} {100}},\ \bibinfo {pages} {245427} (\bibinfo {year} {2019})}\BibitemShut {NoStop}%
\bibitem [{\citenamefont {Hu}, \citenamefont {Liu},\ and\ \citenamefont {Nori}(2012)}]{nori2012}%
  \BibitemOpen
  \bibfield  {author} {\bibinfo {author} {\bibfnamefont {X.}~\bibnamefont {Hu}}, \bibinfo {author} {\bibfnamefont {Y.}~\bibnamefont {Liu}},\ and\ \bibinfo {author} {\bibfnamefont {F.}~\bibnamefont {Nori}},\ }\href@noop {} {\bibfield  {journal} {\bibinfo  {journal} {Phys. Rev. B}\ }\textbf {\bibinfo {volume} {86}},\ \bibinfo {pages} {035314} (\bibinfo {year} {2012})}\BibitemShut {NoStop}%
\bibitem [{\citenamefont {Guo}, \citenamefont {Konstantinov},\ and\ \citenamefont {Jin}(2025)}]{guo2025quantum}%
  \BibitemOpen
  \bibfield  {author} {\bibinfo {author} {\bibfnamefont {W.}~\bibnamefont {Guo}}, \bibinfo {author} {\bibfnamefont {D.}~\bibnamefont {Konstantinov}},\ and\ \bibinfo {author} {\bibfnamefont {D.}~\bibnamefont {Jin}},\ }\bibfield  {title} {\enquote {\bibinfo {title} {Quantum electronics on quantum liquids and solids},}\ }\href@noop {} {\bibfield  {journal} {\bibinfo  {journal} {Progress in Quantum Electronics}\ }\textbf {\bibinfo {volume} {99}},\ \bibinfo {pages} {100552} (\bibinfo {year} {2025})}\BibitemShut {NoStop}%
\bibitem [{\citenamefont {Zavyalov}\ \emph {et~al.}(2005)\citenamefont {Zavyalov}, \citenamefont {Smolyaninov}, \citenamefont {Zotova}, \citenamefont {Borodin},\ and\ \citenamefont {Bogomolov}}]{zavyalov2005}%
  \BibitemOpen
  \bibfield  {author} {\bibinfo {author} {\bibfnamefont {V.}~\bibnamefont {Zavyalov}}, \bibinfo {author} {\bibfnamefont {I.}~\bibnamefont {Smolyaninov}}, \bibinfo {author} {\bibfnamefont {E.}~\bibnamefont {Zotova}}, \bibinfo {author} {\bibfnamefont {A.}~\bibnamefont {Borodin}},\ and\ \bibinfo {author} {\bibfnamefont {S.}~\bibnamefont {Bogomolov}},\ }\href@noop {} {\bibfield  {journal} {\bibinfo  {journal} {J. Low Temp. Phys.}\ }\textbf {\bibinfo {volume} {138}},\ \bibinfo {pages} {415} (\bibinfo {year} {2005})}\BibitemShut {NoStop}%
\bibitem [{\citenamefont {Cole}\ and\ \citenamefont {Cohen}(1969)}]{cole1969}%
  \BibitemOpen
  \bibfield  {author} {\bibinfo {author} {\bibfnamefont {M.}~\bibnamefont {Cole}}\ and\ \bibinfo {author} {\bibfnamefont {M.~H.}\ \bibnamefont {Cohen}},\ }\bibfield  {title} {\enquote {\bibinfo {title} {Image-potential-induced surface bands in insulators},}\ }\href@noop {} {\bibfield  {journal} {\bibinfo  {journal} {Phys. Rev. Lett.}\ }\textbf {\bibinfo {volume} {23}},\ \bibinfo {pages} {1238} (\bibinfo {year} {1969})}\BibitemShut {NoStop}%
\bibitem [{\citenamefont {Cole}(1971)}]{cole1971}%
  \BibitemOpen
  \bibfield  {author} {\bibinfo {author} {\bibfnamefont {M.}~\bibnamefont {Cole}},\ }\bibfield  {title} {\enquote {\bibinfo {title} {Electronic surface states of a dielectric film on a metal substrate},}\ }\href@noop {} {\bibfield  {journal} {\bibinfo  {journal} {Phys. Rev. B}\ }\textbf {\bibinfo {volume} {3}},\ \bibinfo {pages} {4418} (\bibinfo {year} {1971})}\BibitemShut {NoStop}%
\bibitem [{\citenamefont {Leiderer}(1992)}]{leiderer1992}%
  \BibitemOpen
  \bibfield  {author} {\bibinfo {author} {\bibfnamefont {P.}~\bibnamefont {Leiderer}},\ }\bibfield  {title} {\enquote {\bibinfo {title} {Electrons at the surface of quantum systems},}\ }\href@noop {} {\bibfield  {journal} {\bibinfo  {journal} {J. Low Temp. Phys.}\ }\textbf {\bibinfo {volume} {87}},\ \bibinfo {pages} {247} (\bibinfo {year} {1992})}\BibitemShut {NoStop}%
\bibitem [{\citenamefont {Koolstra}, \citenamefont {Yang},\ and\ \citenamefont {Schuster}(2019)}]{koolstra2019}%
  \BibitemOpen
  \bibfield  {author} {\bibinfo {author} {\bibfnamefont {G.}~\bibnamefont {Koolstra}}, \bibinfo {author} {\bibfnamefont {G.}~\bibnamefont {Yang}},\ and\ \bibinfo {author} {\bibfnamefont {D.~I.}\ \bibnamefont {Schuster}},\ }\bibfield  {title} {\enquote {\bibinfo {title} {Coupling a single electron on superfluid helium to a superconducting resonator},}\ }\href@noop {} {\bibfield  {journal} {\bibinfo  {journal} {Nat. Commun.}\ }\textbf {\bibinfo {volume} {10}},\ \bibinfo {pages} {5323} (\bibinfo {year} {2019})}\BibitemShut {NoStop}%
\bibitem [{\citenamefont {Koolstra}\ \emph {et~al.}(2025)\citenamefont {Koolstra}, \citenamefont {Glen}, \citenamefont {Beysengulov}, \citenamefont {Byeon}, \citenamefont {Castoria}, \citenamefont {Sammon}, \citenamefont {Dizdar}, \citenamefont {Wang}, \citenamefont {Schuster}, \citenamefont {Lyon}, \citenamefont {Pollanen},\ and\ \citenamefont {Rees}}]{koolstra2025high}%
  \BibitemOpen
  \bibfield  {author} {\bibinfo {author} {\bibfnamefont {G.}~\bibnamefont {Koolstra}}, \bibinfo {author} {\bibfnamefont {E.}~\bibnamefont {Glen}}, \bibinfo {author} {\bibfnamefont {N.}~\bibnamefont {Beysengulov}}, \bibinfo {author} {\bibfnamefont {H.}~\bibnamefont {Byeon}}, \bibinfo {author} {\bibfnamefont {K.}~\bibnamefont {Castoria}}, \bibinfo {author} {\bibfnamefont {M.}~\bibnamefont {Sammon}}, \bibinfo {author} {\bibfnamefont {B.}~\bibnamefont {Dizdar}}, \bibinfo {author} {\bibfnamefont {C.}~\bibnamefont {Wang}}, \bibinfo {author} {\bibfnamefont {D.}~\bibnamefont {Schuster}}, \bibinfo {author} {\bibfnamefont {S.}~\bibnamefont {Lyon}}, \bibinfo {author} {\bibfnamefont {J.}~\bibnamefont {Pollanen}},\ and\ \bibinfo {author} {\bibfnamefont {D.}~\bibnamefont {Rees}},\ }\bibfield  {title} {\enquote {\bibinfo {title} {High-impedance resonators for strong coupling to an electron on helium},}\ }\href@noop {} {\bibfield  {journal} {\bibinfo  {journal} {Physical Review Applied}\ }\textbf {\bibinfo {volume} {23}},\
  \bibinfo {pages} {024001} (\bibinfo {year} {2025})}\BibitemShut {NoStop}%
\bibitem [{\citenamefont {Yang}\ \emph {et~al.}(2016)\citenamefont {Yang}, \citenamefont {Fragner}, \citenamefont {Koolstra}, \citenamefont {Ocola}, \citenamefont {Czaplewski}, \citenamefont {Schoelkopf},\ and\ \citenamefont {Schuster}}]{yang2016}%
  \BibitemOpen
  \bibfield  {author} {\bibinfo {author} {\bibfnamefont {G.}~\bibnamefont {Yang}}, \bibinfo {author} {\bibfnamefont {A.}~\bibnamefont {Fragner}}, \bibinfo {author} {\bibfnamefont {G.}~\bibnamefont {Koolstra}}, \bibinfo {author} {\bibfnamefont {L.}~\bibnamefont {Ocola}}, \bibinfo {author} {\bibfnamefont {D.~A.}\ \bibnamefont {Czaplewski}}, \bibinfo {author} {\bibfnamefont {R.~J.}\ \bibnamefont {Schoelkopf}},\ and\ \bibinfo {author} {\bibfnamefont {D.~I.}\ \bibnamefont {Schuster}},\ }\bibfield  {title} {\enquote {\bibinfo {title} {Coupling an ensemble of electrons on superfluid helium to a superconducting circuit},}\ }\href@noop {} {\bibfield  {journal} {\bibinfo  {journal} {Phys. Rev. X}\ }\textbf {\bibinfo {volume} {6}},\ \bibinfo {pages} {011031} (\bibinfo {year} {2016})}\BibitemShut {NoStop}%
\bibitem [{\citenamefont {Zhou}\ \emph {et~al.}(2022)\citenamefont {Zhou}, \citenamefont {Koolstra}, \citenamefont {Zhang}, \citenamefont {Yang}, \citenamefont {Han}, \citenamefont {Dizdar}, \citenamefont {Li}, \citenamefont {Ralu}, \citenamefont {Guo}, \citenamefont {Murch}, \citenamefont {Schuster},\ and\ \citenamefont {Jin}}]{zhou2022}%
  \BibitemOpen
  \bibfield  {author} {\bibinfo {author} {\bibfnamefont {X.}~\bibnamefont {Zhou}}, \bibinfo {author} {\bibfnamefont {G.}~\bibnamefont {Koolstra}}, \bibinfo {author} {\bibfnamefont {X.}~\bibnamefont {Zhang}}, \bibinfo {author} {\bibfnamefont {G.}~\bibnamefont {Yang}}, \bibinfo {author} {\bibfnamefont {X.}~\bibnamefont {Han}}, \bibinfo {author} {\bibfnamefont {B.}~\bibnamefont {Dizdar}}, \bibinfo {author} {\bibfnamefont {X.}~\bibnamefont {Li}}, \bibinfo {author} {\bibfnamefont {D.}~\bibnamefont {Ralu}}, \bibinfo {author} {\bibfnamefont {W.}~\bibnamefont {Guo}}, \bibinfo {author} {\bibfnamefont {K.~W.}\ \bibnamefont {Murch}}, \bibinfo {author} {\bibfnamefont {D.~I.}\ \bibnamefont {Schuster}},\ and\ \bibinfo {author} {\bibfnamefont {D.}~\bibnamefont {Jin}},\ }\bibfield  {title} {\enquote {\bibinfo {title} {Single electrons on solid neon as a solid-state qubit platform},}\ }\href@noop {} {\bibfield  {journal} {\bibinfo  {journal} {Nature}\ }\textbf {\bibinfo {volume} {605}},\ \bibinfo {pages} {46--50} (\bibinfo
  {year} {2022})}\BibitemShut {NoStop}%
\bibitem [{\citenamefont {Zhou}\ \emph {et~al.}(2024)\citenamefont {Zhou}, \citenamefont {Li}, \citenamefont {Chen}, \citenamefont {Koolstra}, \citenamefont {Yang}, \citenamefont {Dizdar}, \citenamefont {Huang}, \citenamefont {Wang}, \citenamefont {Han}, \citenamefont {Zhang}, \citenamefont {Schuster},\ and\ \citenamefont {Jin}}]{zhou2024electron}%
  \BibitemOpen
  \bibfield  {author} {\bibinfo {author} {\bibfnamefont {X.}~\bibnamefont {Zhou}}, \bibinfo {author} {\bibfnamefont {X.}~\bibnamefont {Li}}, \bibinfo {author} {\bibfnamefont {Q.}~\bibnamefont {Chen}}, \bibinfo {author} {\bibfnamefont {G.}~\bibnamefont {Koolstra}}, \bibinfo {author} {\bibfnamefont {G.}~\bibnamefont {Yang}}, \bibinfo {author} {\bibfnamefont {B.}~\bibnamefont {Dizdar}}, \bibinfo {author} {\bibfnamefont {Y.}~\bibnamefont {Huang}}, \bibinfo {author} {\bibfnamefont {C.~S.}\ \bibnamefont {Wang}}, \bibinfo {author} {\bibfnamefont {X.}~\bibnamefont {Han}}, \bibinfo {author} {\bibfnamefont {X.}~\bibnamefont {Zhang}}, \bibinfo {author} {\bibfnamefont {D.}~\bibnamefont {Schuster}},\ and\ \bibinfo {author} {\bibfnamefont {D.}~\bibnamefont {Jin}},\ }\bibfield  {title} {\enquote {\bibinfo {title} {Electron charge qubit with 0.1 millisecond coherence time},}\ }\href@noop {} {\bibfield  {journal} {\bibinfo  {journal} {Nature Physics}\ }\textbf {\bibinfo {volume} {20}},\ \bibinfo {pages} {116--122} (\bibinfo
  {year} {2024})}\BibitemShut {NoStop}%
\bibitem [{\citenamefont {Heinrich}\ \emph {et~al.}(2021)\citenamefont {Heinrich}, \citenamefont {Oliver}, \citenamefont {Vandersypen}, \citenamefont {Ardavan}, \citenamefont {Sessoli}, \citenamefont {Loss}, \citenamefont {Jayich}, \citenamefont {Fernandez-Rossier}, \citenamefont {Laucht},\ and\ \citenamefont {Morello}}]{heinrich2021}%
  \BibitemOpen
  \bibfield  {author} {\bibinfo {author} {\bibfnamefont {A.~J.}\ \bibnamefont {Heinrich}}, \bibinfo {author} {\bibfnamefont {W.~D.}\ \bibnamefont {Oliver}}, \bibinfo {author} {\bibfnamefont {L.~M.}\ \bibnamefont {Vandersypen}}, \bibinfo {author} {\bibfnamefont {A.}~\bibnamefont {Ardavan}}, \bibinfo {author} {\bibfnamefont {R.}~\bibnamefont {Sessoli}}, \bibinfo {author} {\bibfnamefont {D.}~\bibnamefont {Loss}}, \bibinfo {author} {\bibfnamefont {A.~B.}\ \bibnamefont {Jayich}}, \bibinfo {author} {\bibfnamefont {J.}~\bibnamefont {Fernandez-Rossier}}, \bibinfo {author} {\bibfnamefont {A.}~\bibnamefont {Laucht}},\ and\ \bibinfo {author} {\bibfnamefont {A.}~\bibnamefont {Morello}},\ }\href@noop {} {\bibfield  {journal} {\bibinfo  {journal} {Nat. Nanotechnol.}\ }\textbf {\bibinfo {volume} {16}},\ \bibinfo {pages} {1318} (\bibinfo {year} {2021})}\BibitemShut {NoStop}%
\bibitem [{\citenamefont {Kanai}, \citenamefont {Jin},\ and\ \citenamefont {Guo}(2024)}]{kanai2024single}%
  \BibitemOpen
  \bibfield  {author} {\bibinfo {author} {\bibfnamefont {T.}~\bibnamefont {Kanai}}, \bibinfo {author} {\bibfnamefont {D.}~\bibnamefont {Jin}},\ and\ \bibinfo {author} {\bibfnamefont {W.}~\bibnamefont {Guo}},\ }\bibfield  {title} {\enquote {\bibinfo {title} {Single-electron qubits based on quantum ring states on solid neon surface},}\ }\href@noop {} {\bibfield  {journal} {\bibinfo  {journal} {Physical Review Letters}\ }\textbf {\bibinfo {volume} {132}},\ \bibinfo {pages} {250603} (\bibinfo {year} {2024})}\BibitemShut {NoStop}%
\bibitem [{\citenamefont {Zheng}, \citenamefont {Song},\ and\ \citenamefont {Murch}(2025)}]{zheng2025surface}%
  \BibitemOpen
  \bibfield  {author} {\bibinfo {author} {\bibfnamefont {K.}~\bibnamefont {Zheng}}, \bibinfo {author} {\bibfnamefont {X.}~\bibnamefont {Song}},\ and\ \bibinfo {author} {\bibfnamefont {K.~W.}\ \bibnamefont {Murch}},\ }\bibfield  {title} {\enquote {\bibinfo {title} {Surface morphology assisted trapping of strongly coupled electron-on-neon charge states},}\ }\href@noop {} {\bibfield  {journal} {\bibinfo  {journal} {arXiv preprint arXiv:2503.01847}\ } (\bibinfo {year} {2025})}\BibitemShut {NoStop}%
\bibitem [{\citenamefont {Andrei}(2012)}]{andrei2012two}%
  \BibitemOpen
  \bibfield  {author} {\bibinfo {author} {\bibfnamefont {E.~Y.}\ \bibnamefont {Andrei}},\ }\bibfield  {title} {\enquote {\bibinfo {title} {Two-dimensional electron systems: on helium and other cryogenic substrates},}\ }\href@noop {} {\bibfield  {journal} {\bibinfo  {journal} {Springer Science \& Business Media}\ }\textbf {\bibinfo {volume} {19}} (\bibinfo {year} {2012})}\BibitemShut {NoStop}%
\bibitem [{\citenamefont {Bethe}\ and\ \citenamefont {Jackiw}(2018)}]{bethe2018}%
  \BibitemOpen
  \bibfield  {author} {\bibinfo {author} {\bibfnamefont {H.~A.}\ \bibnamefont {Bethe}}\ and\ \bibinfo {author} {\bibfnamefont {R.}~\bibnamefont {Jackiw}},\ }\bibfield  {title} {\enquote {\bibinfo {title} {Intermediate quantum mechanics: Third edition (frontiers in physics)},}\ }\href@noop {} {\bibfield  {journal} {\bibinfo  {journal} {CRC Press}\ } (\bibinfo {year} {2018})}\BibitemShut {NoStop}%
\bibitem [{\citenamefont {Traficante}(1991)}]{traficante1991}%
  \BibitemOpen
  \bibfield  {author} {\bibinfo {author} {\bibfnamefont {D.~D.}\ \bibnamefont {Traficante}},\ }\bibfield  {title} {\enquote {\bibinfo {title} {Relaxation. can \text{T}$_{2}$, be longer than \text{T}$_{1}$?}}\ }\href@noop {} {\bibfield  {journal} {\bibinfo  {journal} {Concepts Magn Reson}\ }\textbf {\bibinfo {volume} {3}},\ \bibinfo {pages} {171} (\bibinfo {year} {1991})}\BibitemShut {NoStop}%
\bibitem [{\citenamefont {Baranov}\ \emph {et~al.}(2017)\citenamefont {Baranov}, \citenamefont {Savelev}, \citenamefont {Li}, \citenamefont {Krasnok},\ and\ \citenamefont {Alu}}]{baranov2017}%
  \BibitemOpen
  \bibfield  {author} {\bibinfo {author} {\bibfnamefont {D.~G.}\ \bibnamefont {Baranov}}, \bibinfo {author} {\bibfnamefont {R.~S.}\ \bibnamefont {Savelev}}, \bibinfo {author} {\bibfnamefont {S.~V.}\ \bibnamefont {Li}}, \bibinfo {author} {\bibfnamefont {A.~E.}\ \bibnamefont {Krasnok}},\ and\ \bibinfo {author} {\bibfnamefont {A.}~\bibnamefont {Alu}},\ }\bibfield  {title} {\enquote {\bibinfo {title} {Modifying magnetic dipole spontaneous emission with nanophotonic structures},}\ }\href@noop {} {\bibfield  {journal} {\bibinfo  {journal} {Laser Photonics Rev.}\ }\textbf {\bibinfo {volume} {11}},\ \bibinfo {pages} {1600268} (\bibinfo {year} {2017})}\BibitemShut {NoStop}%
\bibitem [{\citenamefont {Reed}\ \emph {et~al.}(2010)\citenamefont {Reed}, \citenamefont {Johnson}, \citenamefont {Houck}, \citenamefont {DiCarlo}, \citenamefont {Chow}, \citenamefont {Schuster}, \citenamefont {Frunzio},\ and\ \citenamefont {Schoelkopf}}]{reed2010fast}%
  \BibitemOpen
  \bibfield  {author} {\bibinfo {author} {\bibfnamefont {M.~D.}\ \bibnamefont {Reed}}, \bibinfo {author} {\bibfnamefont {B.~R.}\ \bibnamefont {Johnson}}, \bibinfo {author} {\bibfnamefont {A.~A.}\ \bibnamefont {Houck}}, \bibinfo {author} {\bibfnamefont {L.}~\bibnamefont {DiCarlo}}, \bibinfo {author} {\bibfnamefont {J.~M.}\ \bibnamefont {Chow}}, \bibinfo {author} {\bibfnamefont {D.~I.}\ \bibnamefont {Schuster}}, \bibinfo {author} {\bibfnamefont {L.}~\bibnamefont {Frunzio}},\ and\ \bibinfo {author} {\bibfnamefont {R.~J.}\ \bibnamefont {Schoelkopf}},\ }\bibfield  {title} {\enquote {\bibinfo {title} {Fast reset and suppressing spontaneous emission of a superconducting qubit},}\ }\href@noop {} {\bibfield  {journal} {\bibinfo  {journal} {Applied Physics Letters}\ }\textbf {\bibinfo {volume} {96}} (\bibinfo {year} {2010})}\BibitemShut {NoStop}%
\bibitem [{\citenamefont {O'Neil}(2013)}]{o2013merck}%
  \BibitemOpen
  \bibfield  {author} {\bibinfo {author} {\bibfnamefont {M.~J.}\ \bibnamefont {O'Neil}},\ }\bibfield  {title} {\enquote {\bibinfo {title} {The merck index: an encyclopedia of chemicals, drugs, and biologicals},}\ }\href@noop {} {\bibfield  {journal} {\bibinfo  {journal} {RSc Publishing}\ } (\bibinfo {year} {2013})}\BibitemShut {NoStop}%
\bibitem [{\citenamefont {Balzer}, \citenamefont {Kupperman},\ and\ \citenamefont {Simmons}(1971)}]{balzer1971}%
  \BibitemOpen
  \bibfield  {author} {\bibinfo {author} {\bibfnamefont {R.}~\bibnamefont {Balzer}}, \bibinfo {author} {\bibfnamefont {D.~S.}\ \bibnamefont {Kupperman}},\ and\ \bibinfo {author} {\bibfnamefont {R.~O.}\ \bibnamefont {Simmons}},\ }\bibfield  {title} {\enquote {\bibinfo {title} {Velocities of sound in polycrystalline neon},}\ }\href@noop {} {\bibfield  {journal} {\bibinfo  {journal} {Phys. Rev. B}\ }\textbf {\bibinfo {volume} {4}},\ \bibinfo {pages} {3636} (\bibinfo {year} {1971})}\BibitemShut {NoStop}%
\bibitem [{\citenamefont {Schuster}\ \emph {et~al.}(2010)\citenamefont {Schuster}, \citenamefont {Fragner}, \citenamefont {Dykman}, \citenamefont {Lyon},\ and\ \citenamefont {Schoelkopf}}]{schuster2010}%
  \BibitemOpen
  \bibfield  {author} {\bibinfo {author} {\bibfnamefont {D.~I.}\ \bibnamefont {Schuster}}, \bibinfo {author} {\bibfnamefont {A.}~\bibnamefont {Fragner}}, \bibinfo {author} {\bibfnamefont {M.~I.}\ \bibnamefont {Dykman}}, \bibinfo {author} {\bibfnamefont {S.~A.}\ \bibnamefont {Lyon}},\ and\ \bibinfo {author} {\bibfnamefont {R.~J.}\ \bibnamefont {Schoelkopf}},\ }\bibfield  {title} {\enquote {\bibinfo {title} {Proposal for manipulating and detecting spin and orbital states of trapped electrons on helium using cavity quantum electrodynamics},}\ }\href@noop {} {\bibfield  {journal} {\bibinfo  {journal} {Phys. Rev. Lett.}\ }\textbf {\bibinfo {volume} {105}},\ \bibinfo {pages} {040503} (\bibinfo {year} {2010})}\BibitemShut {NoStop}%
\bibitem [{\citenamefont {Dykman}, \citenamefont {Platzman},\ and\ \citenamefont {Seddighrad}(2003)}]{dykman2003}%
  \BibitemOpen
  \bibfield  {author} {\bibinfo {author} {\bibfnamefont {M.~I.}\ \bibnamefont {Dykman}}, \bibinfo {author} {\bibfnamefont {P.~M.}\ \bibnamefont {Platzman}},\ and\ \bibinfo {author} {\bibfnamefont {P.}~\bibnamefont {Seddighrad}},\ }\bibfield  {title} {\enquote {\bibinfo {title} {Qubits with electrons on liquid helium},}\ }\href@noop {} {\bibfield  {journal} {\bibinfo  {journal} {Phys. Rev. B}\ }\textbf {\bibinfo {volume} {67}},\ \bibinfo {pages} {155402} (\bibinfo {year} {2003})}\BibitemShut {NoStop}%
\bibitem [{\citenamefont {Saitoh}(1977)}]{saitoh1977}%
  \BibitemOpen
  \bibfield  {author} {\bibinfo {author} {\bibfnamefont {M.}~\bibnamefont {Saitoh}},\ }\href@noop {} {\bibfield  {journal} {\bibinfo  {journal} {J. Phys. Soc. Jpn.}\ }\textbf {\bibinfo {volume} {42}},\ \bibinfo {pages} {201} (\bibinfo {year} {1977})}\BibitemShut {NoStop}%
\bibitem [{\citenamefont {Monarkha}\ and\ \citenamefont {Shikin}(1982)}]{monarkha1982}%
  \BibitemOpen
  \bibfield  {author} {\bibinfo {author} {\bibfnamefont {Y.}~\bibnamefont {Monarkha}}\ and\ \bibinfo {author} {\bibfnamefont {V.}~\bibnamefont {Shikin}},\ }\bibfield  {title} {\enquote {\bibinfo {title} {Low-dimensional electron systems at liquid helium surface},}\ }\href@noop {} {\bibfield  {journal} {\bibinfo  {journal} {Fiz. Nizk. Temp.}\ }\textbf {\bibinfo {volume} {8}},\ \bibinfo {pages} {563} (\bibinfo {year} {1982})}\BibitemShut {NoStop}%
\bibitem [{\citenamefont {Chen}\ \emph {et~al.}(2022)\citenamefont {Chen}, \citenamefont {I.Martin}, \citenamefont {Jiang},\ and\ \citenamefont {Jin}}]{chen2022}%
  \BibitemOpen
  \bibfield  {author} {\bibinfo {author} {\bibfnamefont {Q.}~\bibnamefont {Chen}}, \bibinfo {author} {\bibnamefont {I.Martin}}, \bibinfo {author} {\bibfnamefont {L.}~\bibnamefont {Jiang}},\ and\ \bibinfo {author} {\bibfnamefont {D.}~\bibnamefont {Jin}},\ }\href@noop {} {\bibfield  {journal} {\bibinfo  {journal} {Quantum Sci. and Technol.}\ }\textbf {\bibinfo {volume} {7}},\ \bibinfo {pages} {045016} (\bibinfo {year} {2022})}\BibitemShut {NoStop}%
\bibitem [{\citenamefont {McSkimin}\ and\ \citenamefont {Jr.}(1964)}]{mcskimin1964}%
  \BibitemOpen
  \bibfield  {author} {\bibinfo {author} {\bibfnamefont {H.~J.}\ \bibnamefont {McSkimin}}\ and\ \bibinfo {author} {\bibfnamefont {P.~A.}\ \bibnamefont {Jr.}},\ }\bibfield  {title} {\enquote {\bibinfo {title} {Elastic moduli of silicon vs hydrostatic pressure at 25.0 c and- 195.8 c.}}\ }\href@noop {} {\bibfield  {journal} {\bibinfo  {journal} {J. Appl. Phys.}\ }\textbf {\bibinfo {volume} {35}},\ \bibinfo {pages} {2161--2165} (\bibinfo {year} {1964})}\BibitemShut {NoStop}%
\bibitem [{\citenamefont {Danil'Chenko}, \citenamefont {Poroshin},\ and\ \citenamefont {Sarbei.}(1979)}]{sarbei1979}%
  \BibitemOpen
  \bibfield  {author} {\bibinfo {author} {\bibfnamefont {B.}~\bibnamefont {Danil'Chenko}}, \bibinfo {author} {\bibfnamefont {V.}~\bibnamefont {Poroshin}},\ and\ \bibinfo {author} {\bibfnamefont {O.}~\bibnamefont {Sarbei.}},\ }\bibfield  {title} {\enquote {\bibinfo {title} {An observation of second sound in sapphire},}\ }\href@noop {} {\bibfield  {journal} {\bibinfo  {journal} {JETP Lett}\ }\textbf {\bibinfo {volume} {30}},\ \bibinfo {pages} {215--218} (\bibinfo {year} {1979})}\BibitemShut {NoStop}%
\end{thebibliography}%

\end{document}